\documentclass{ws-procs975x65}

\begin{document}
\vskip -2.0cm
\title{Parameter space study of magnetohydrodynamic flows 
around magnetized compact objects}

\author{Santabrata Das$^1$ and Sandip K. Chakrabarti$^{2,3}$}

\address{$^1$ARCSEC, Sejong University,
%         98 Gunja-Dong, Gwangjin-Gu,
         Seoul 143-747, Korea.
E-mail: sbdas@canopus.cnu.ac.kr\\}

\address{$^2$ SNBNCBS,%S. N. Bose National centre for Basic Sciences,
  JD-Block, Sector III,
  Salt Lake, Kolkata-98, India. E-mail:chakraba@bose.res.in\\
  $^3$ Centre for Space Physics, Chalantika 43, Garia Station Rd.,
  Kolkata 700084, India.}

\begin{abstract}
We solve the magnetohydrodynamic (MHD) equations governing axisymmetric
flows around neutron stars and black holes and found all possible
solution topologies for adiabatic accretion. We divide the parameter 
space spanned by the conserved energy and angular momentum of the flow in terms
of the flow topologies. We also study the possibility of the formation of
the MHD shock waves.
\end{abstract}


\bodymatter

\section{Introduction}

In recent years, the study of magnetohydrodynamic (MHD) accretion
flow around compact objects has become very important since
the magnetic field in ubiquitous in the universe and it should play a part in the 
black hole accretion, especially close to the inner disk. The pioneering works of
Mestel (1967)\cite{m67} and Weber \& Davis (1967) \cite{wd67} 
etc. were further extended by  Chakrabarti (1990) \cite{c90} for rotating compact stars
and obtained a few global solutions for accretion and winds. 
The study of MHD flow in Kerr geometry is also carried out by 
Takahashi et al. (1990) \cite{tetal90}. They discussed the properties of smooth 
trans-Alf\'venic flows and focused on the possible ways to 
extract energies out of the BH. Nitta et al. (1991) \cite{netal91}
presented an analytical solution of general relativistic
Grad-Safranov equation around a rotating BH and obtained
few solution topologies. So far, a fully self-consistent
study of the global solutions around a compact star in terms of 
flow parameters was not explored. In this paper, we wish to
provide a complete study of the trans-magnetosonic accretion flow 
around compact objects in presence of both the radial and toroidal 
magnetic fields using Paczy\'nski-Wiita \cite{pw80} pseudo-Newtonian potential. 
We discuss how the nature of the solution topologies
depend on the parameter space spanned by the energy (${\cal E}$) and the angular
momentum ($L$) of the flow. We identify eighteen
types of solution topologies and divide the parameter space accordingly.
We also find that the standing magneto-hydrodynamic shock wave forms
which could give rise to the high energy particles (~ few MeV) through the
shock acceleration mechanism of the electrons. These accelerated particles 
produce power-law synchrotron radiations which could explain some of the 
spectral properties of BH candidates\cite{cm06}. Moreover, 
soft photons are energized by the hot electrons of post-shock
flow through inverse Comptonization. In this paper, we 
provide all possible solutions of MHD flow with or without shocks 
and classify the parameter space according to the nature of the 
flow solution.

\def\figsubcap#1{\par\noindent\centering\footnotesize(#1)}
\begin{figure}[b]%
\vskip -0.35 in
\begin{center}
  \parbox{2.1in}{\epsfig{figure=fig4.eps,width=2in}\figsubcap{a}}
%  \hspace*{4pt}
\hskip 0.5cm
  \parbox{2.1in}{\epsfig{figure=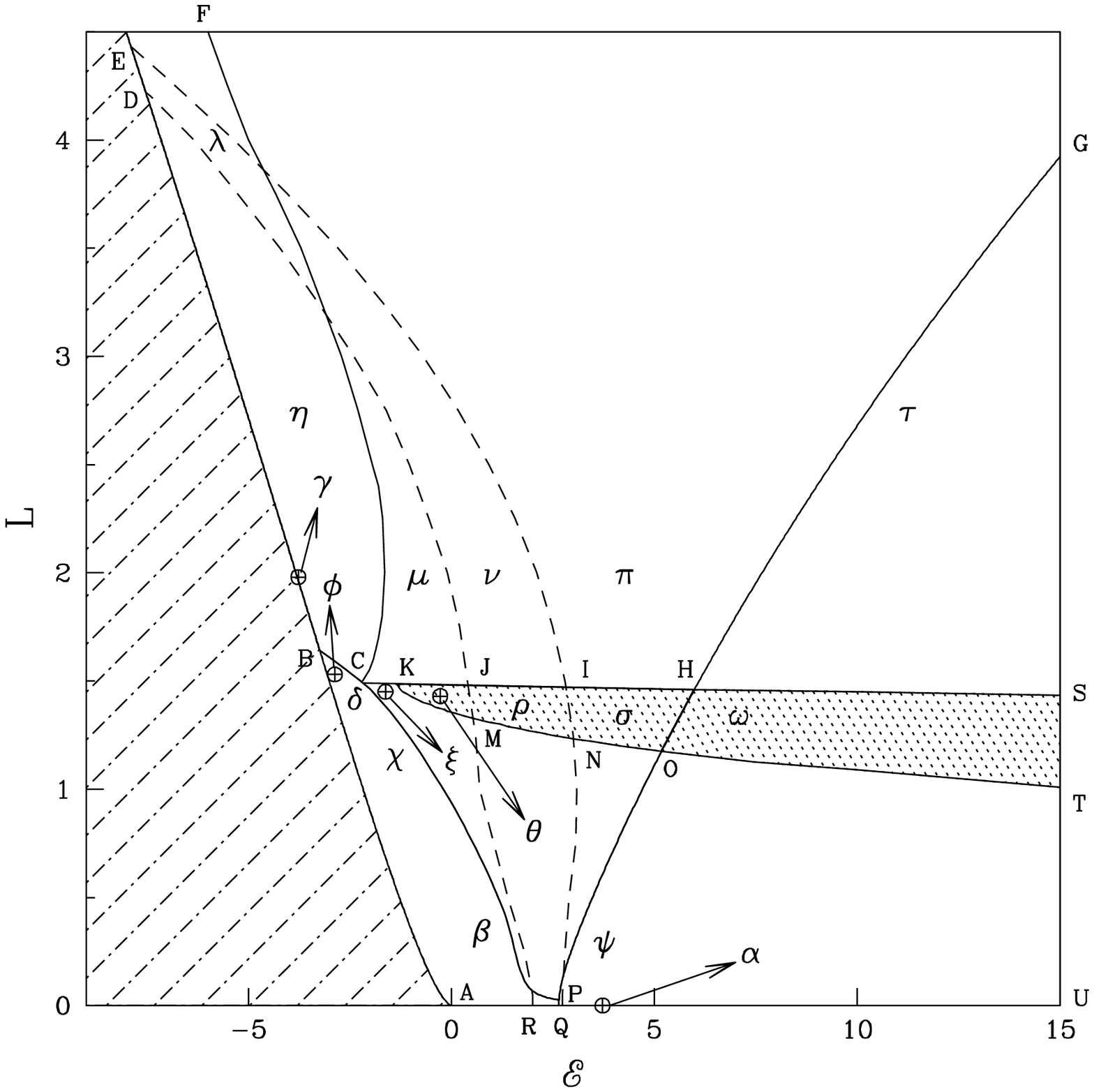,width=2.1in}\figsubcap{b}}
\hskip 0.5cm
  \parbox{2.1in}{\epsfig{figure=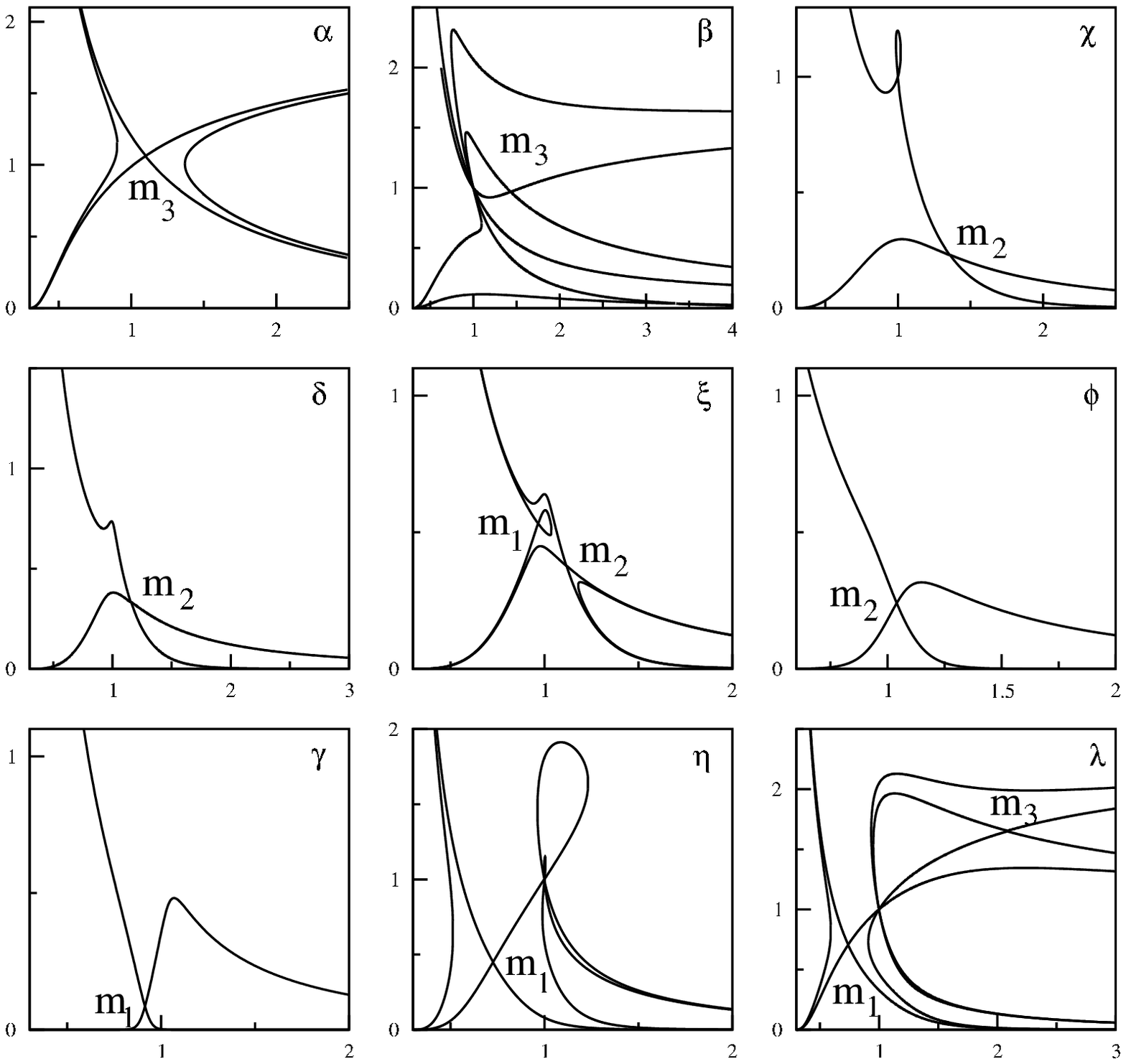,width=2.0in}\figsubcap{c}}
\hskip 0.5cm
  \parbox{2.6in}{\epsfig{figure=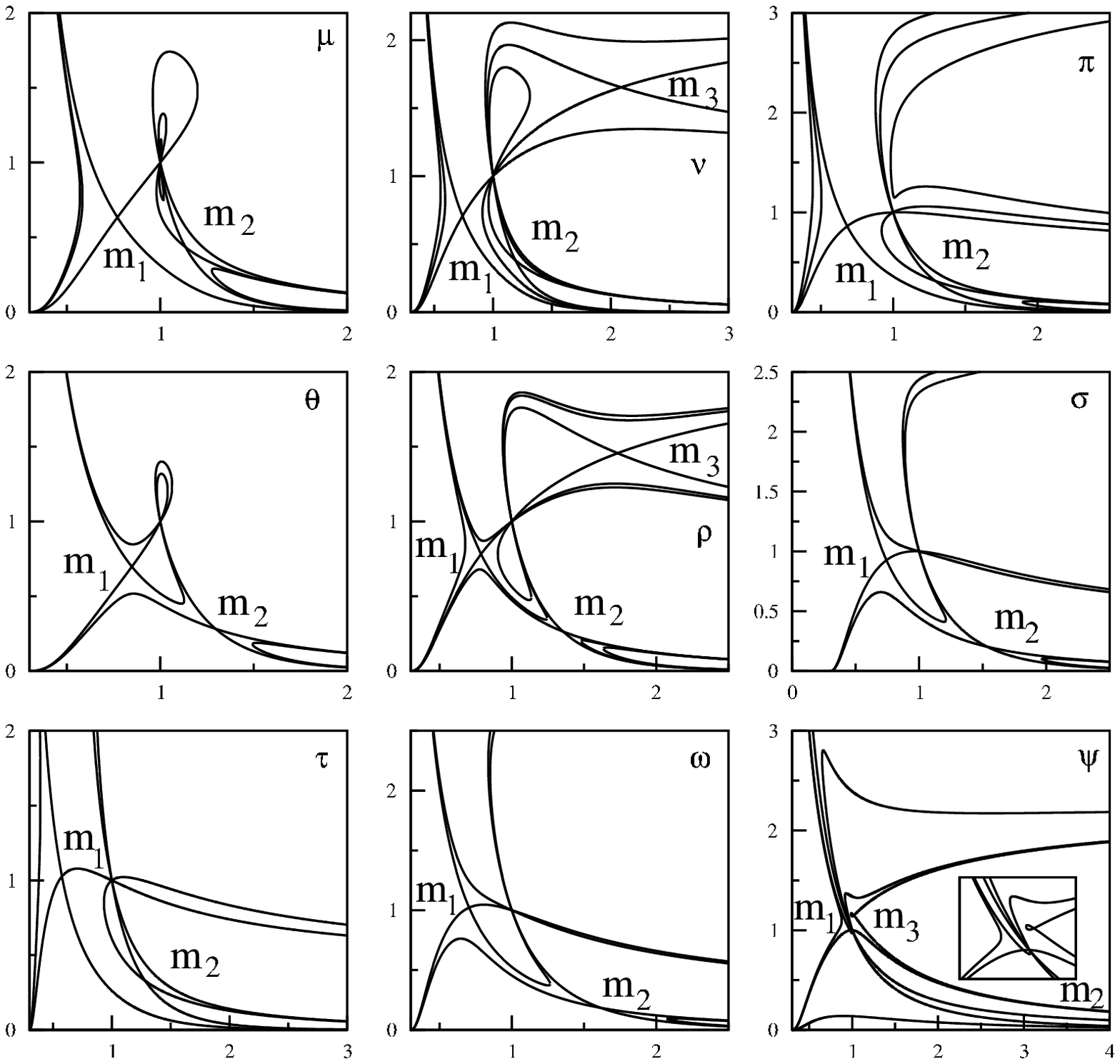,width=2.0in}\figsubcap{d}}
  \caption{(a) A complete solution topology with standing 
               magnetohydrodynamic shock wave\cite{dc07}. 
               (b) Division of the parameter space according to the nature of solution topologies\cite{dc07}.
               (c-d)
               MHD flow solutions for different (${\cal E}, L$) pair. Radial
               distance is plotted along horizontal axis while radial
               velocity is plotted along vertical axis. Greek
               alphabets mark various types of solutions drawn with 
               parameters from different regions marked Fig. 1b\cite{dc07}.}%
  \label{fig1.2}
\end{center}
\end{figure}

\section{Behaviour of MHD shocks, flow solutions and parameter space}

An example of complete MHD solution with standing, slow-magnetosonic shock 
is presented in Fig. 1a. We consider the Alfv{\'e}n velocity and the radius  to be
$\vartheta_a=10^{10}$ cm s$^{-1}$ and $r_a= 10^7$cm respectively. The
central mass was chosen to be $10M_\odot$ for illustration purpose. The
flow parameters are (${\cal E}$, $L$) = $(1.5,\ 1.45)$. The shock location 
is predicted at $r_s=1.153$ and denoted by the vertical dashed line.
The solution towards the BH is depicted by a single-arrowed curve while
the sub-Alfv{\'enic} double-arrowed curve is appropriate for a 
shock free neutron star solution. The dash-dotted vertical line with 
triple-arrow represents a typical shock transition at the boundary of the
neutron star. The hot and dense post-shock flow is the most important 
region since it is responsible for generating the hard X rays.

In Fig. 1b, we draw the parameter space in ${\cal E}$-$L$ plane and divide 
it into various regions according to the nature of the solution topologies. 
We obtain the curve $ADE$ considering the special case 
where the radial velocity at all the magnetosonic points vanishes. 
The region shaded with dot-dashed lines is forbidden for the flow solution. 
The curve PG is obtained when flow velocity is identical to the Alfv{\'e}n 
velocity at the magnetosonic point. The regions bounded by PBEGHP, ABCFGUA,
RDEQR and QGUQ possess Bondi-like slow, Rotational-slow, Rotational-fast and
Bondi-like fast magnetosonic points respectively. 
The parameters from the region shaded with dots (SKTS) allow
magnetosonic shock transition. The region surrounded by PCKOTUQP 
may produce non-steady shock since shock conditions are not
satisfied but entropy at the inner sonic point is higher than the 
outer sonic point.

We further sub-divide the parameter space according to the solution topologies.
We identified eighteen distinct types of solutions as shown in Fig. 1(c-d).
The accretion solutions for BH and neutron star differs only through the inner
boundary conditions. Therefore, for a set of flow parameters, one needs to
assign the proper boundary conditions to decide whether the solution is for
BH or for neutron star. We marked each solution by a Greek alphabet $\alpha, \beta, ..$ 
etc which denote the parameters in ${\cal E}$-$L$ plane (Fig. 1b) 
for which the solutions are drawn. In Fig. 1(c-d), the Bondi-like (slow/fast), 
Rotational slow and Rotational fast magnetosonic points are denoted 
by $m_1$, $m_2$, and $m_3$ respectively. For $L=0$, the solution reduce to the Bondi flow solution.

\section{Concluding Remarks}

We identified all possible solutions including those containing standing shocks
in a MHD accretion flow around compact objects and divided the 
parameter space according to the nature of flow solutions. 
We identified another region where the
flow may exhibit non-steady shocks very similar to those in a hydrodynamic flow.
The post-shock flow inverse Comptonizes soft photons either from the cooled Keplerian
disk or from the synchrotron radiation and re-emits them as hard X-rays\cite{cm06}. 
In addition, high frequency QPOs of hard X-ray could be obtained since MHD shocks 
form closer to the BH horizon. 

\vskip 0.3cm
\noindent Acknowledgments:
This work is partly supported by a project (Grant No. SP/S2/K-15/2001) funded by DST, India. 
SD is thankful for financial support to KOSEF through ARCSEC, Korea.

{}


\begin{thebibliography}{9}

\bibitem {m67} Mestel, L., 1967, {\it Plasma Astrophysics}, ed. Sturrock, P. A., Academic Press, New York.
\bibitem{wd67}Weber, E. J. \& Davis, L. Jr., 1967, ApJ 148, 217.
\bibitem{c90}Chakrabarti, S. K., 1990,  MNRAS 246, 134.
\bibitem{tetal90}Takahashi, M., Nitta, S., Tatematsu, Y.  \& Tomimatsu, A., 1990, ApJ 363, 206.
\bibitem{netal91}Nitta, S., Takahashi, M. \& Tomimatsu, A., 1991,  Phys. Rev. D 44, 2295.
\bibitem{pw80}Paczy\'{n}ski B., Wiita P.J., 1980,  A \& A 88, 23.
\bibitem{cm06}Chakrabarti, S. K. and S. Mandal, 2006, ApJ, 642, L49.
\bibitem{dc07}Das, S., Chakrabarti, S. K., 2007, MNRAS, 374, 729.
\end{thebibliography}
\end{document}